\documentclass[conference]{IEEEtran}
\IEEEoverridecommandlockouts
\usepackage{cite}
\usepackage{amsmath,amssymb,amsfonts}
\usepackage{algorithmic}
\usepackage{graphicx}
\usepackage{textcomp}
\usepackage{xcolor}
\usepackage{soul}
\usepackage{braket}
\usepackage[a4paper, total={184mm,239mm}]{geometry}

\def\ddefloop#1{\ifx\ddefloop#1\else\ddef{#1}\expandafter\ddefloop\fi}
\def\ddef#1{\expandafter\def\csname bb#1\endcsname{\ensuremath{\mathbb{#1}}}}
\ddefloop ABCDEFGHIJKLMNOPQRSTUVWXYZ\ddefloop
\def\ddef#1{\expandafter\def\csname c#1\endcsname{\ensuremath{\mathcal{#1}}}}
\ddefloop ABCDEFGHIJKLMNOPQRSTUVWXYZ\ddefloop
\def\ddef#1{\expandafter\def\csname v#1\endcsname{\ensuremath{\boldsymbol{#1}}}}
\ddefloop ABCDEFGHIJKLMNOPQRSTUVWXYZabcdefghijklmnopqrstuvwxyz\ddefloop
\def\ddef#1{\expandafter\def\csname v#1\endcsname{\ensuremath{\boldsymbol{\csname #1\endcsname}}}}
\ddefloop {alpha}{beta}{gamma}{delta}{epsilon}{varepsilon}{zeta}{eta}{theta}{vartheta}{iota}{kappa}{lambda}{mu}{nu}{xi}{pi}{varpi}{rho}{varrho}{sigma}{varsigma}{tau}{upsilon}{phi}{varphi}{chi}{psi}{omega}{Gamma}{Delta}{Theta}{Lambda}{Xi}{Pi}{Sigma}{varSigma}{Upsilon}{Phi}{Psi}{Omega}{ell}\ddefloop

\def\BibTeX{{\rm B\kern-.05em{\sc i\kern-.025em b}\kern-.08em
    T\kern-.1667em\lower.7ex\hbox{E}\kern-.125emX}}
\begin{document}

\title{Scalable Variational Quantum Circuits for Autoencoder-based Drug Discovery
\vspace{-4mm}
}

\author{\IEEEauthorblockN{Junde Li and Swaroop Ghosh}
\IEEEauthorblockA{\textit{Department of Computer Science and Engineering} \\
\textit{The Pennsylvania State University}\\
\{jul1512, szg212\}@psu.edu}

}
\IEEEaftertitletext{\vspace{-2\baselineskip}}
\maketitle

\begin{abstract}

The de novo design of drug molecules is recognized as a time-consuming and costly process, and computational approaches have been applied in each stage of the drug discovery pipeline. Variational autoencoder is one of the computer-aided design methods which explores the chemical space based on existing molecular dataset. Quantum machine learning has emerged as an atypical learning method that may speed up some classical learning tasks because of its strong expressive power. However, near-term quantum computers suffer from limited number of qubits which hinders the representation learning in high dimensional spaces. We present a scalable quantum generative autoencoder (SQ-VAE) for simultaneously reconstructing and sampling drug molecules, and a corresponding vanilla variant (SQ-AE) for better reconstruction. The architectural strategies in hybrid quantum classical networks such as, adjustable quantum layer depth, heterogeneous learning rates, and patched quantum circuits are proposed to learn high dimensional dataset such as, ligand-targeted drugs. Extensive experimental results are reported for different dimensions including 8x8 and 32x32 after choosing suitable architectural strategies. The performance of quantum generative autoencoder is compared with the corresponding classical counterpart throughout all experiments. The results show that quantum computing advantages can be achieved for normalized low-dimension molecules, and that high-dimension molecules generated from quantum generative autoencoders have better drug properties within the same learning period.

\end{abstract}


\section{Introduction}

Traditional drug discovery and development pipeline, from concept and market, takes 10-18 years and costs billions of dollars \cite{myers2001drug}.
As the productivity for small-molecule drugs declines, the drug discovery trend shifts toward biologic drugs which predominantly target disease-selective cellular ligands (i.e., molecules that bind to certain atoms to form biological complexes) \cite{li2018ligandomics}. Ligand-targeted drugs exhibiting little affinity for health cells but high affinity for pathologic cells can achieve required therapeutic potency with minimal toxicity \cite{srinivasarao2017ligand}. Ligands would interact with protein pockets (associated with a disease) in the binding sites if they are structurally complementary. One can find such ligands by searching through all viable chemical compounds and molecules adhering to a given set of construction principles and boundary conditions called the \emph{chemical space}. Navigating this impractically large chemical space falls within the field of \textit{de novo} drug discovery \cite{reymond2012exploring}.

\begin{figure}
\centering
\includegraphics[width=8cm]{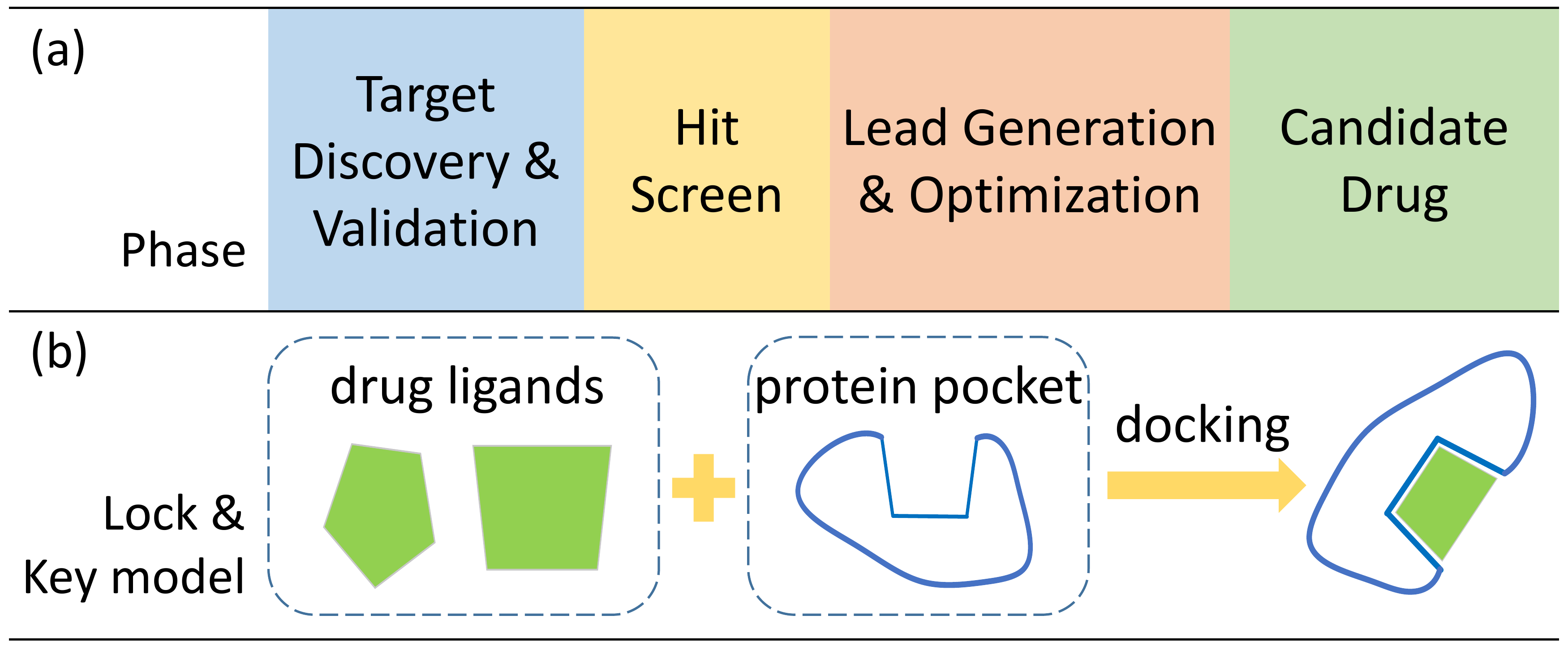}

\vspace{-3mm}
\caption{(a) The drug discovery pipeline in 4 general phases (not including development, clinical trials and launch phases); (b) the lock and key model showing the binding process between protein pocket and ligand.}
\label{drug_discovery}
\vspace{-3mm}
\end{figure}

\begin{figure*}
\centering
\includegraphics[width=17.5cm]{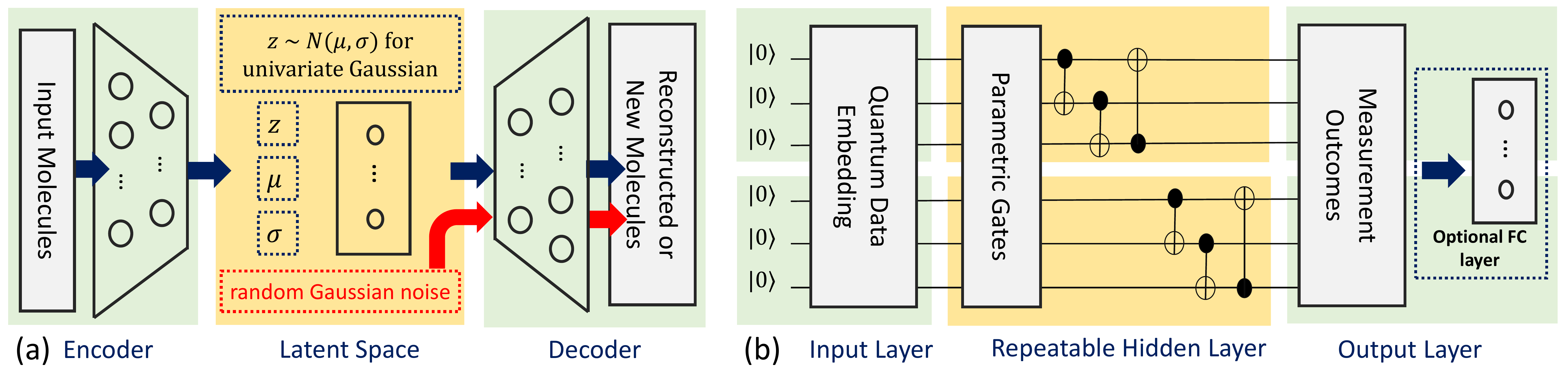}

\vspace{-3mm}
\caption{(a) The classical VAE architecture for drug molecule discovery includes a decoder and an encoder which outputs Gaussian parameters for approximating the latent space distribution (the only part that AE does not involve) from which random noises are sampled for generating new molecules through the decoder network (the generation or sampling process is denoted in red box and arrows); (b) a patched variational quantum circuit (each patch has three qubits) serves as the quantum encoder/decoder network with corresponding quantum input, hidden and output layers. A hybrid VAE/AE is realized by connecting the final fully connected (FC) layer with the preceding measurement outcome vector.}
\label{architecture}
\vspace{-3mm}
\end{figure*}

Computational approaches have been broadly applied to nearly every drug discovery phases, especially molecule design with desirable properties \cite{ekins2019} which corresponds to the lead generation phase shown in Fig. \ref{drug_discovery}(a). Generative models such as, variational autoencoders (VAEs) \cite{vae} and generative adversarial networks (GANs) \cite{gan} are adopted to generate a plethora of drug candidates (in string-based and graph-based \cite{kusner2017grammar, molgan} representations) for further screening by adopting the lock and key model as shown in Fig. \ref{drug_discovery}(b). 
The set-level adversarial loss of GAN is calculated by comparing with the set of training samples; while generative VAEs add extra latent variables from vanilla autoencoders (AEs) and thus support instance-level reconstruction. Such reconstruction is helpful for finding a ligand match given a receptor protein \cite{parmar2021dual, li2021drug}, by going through the molecular docking process as illustrated in Fig. \ref{drug_discovery}(b). AEs support more accurate reconstruction for the lack of latent variables but do not support sampling new ligand molecules.
Therefore, both AEs and VAEs are adopted for reconstructing graph-based molecules but only VAEs for sampling new ligand molecules through learning the latent space distribution as shown in Fig. \ref{architecture}(a).

Quantum computing has shown unique advantages such as, strong expressive power across a variety of learning and optimization applications e.g., detection, classification and discovery \cite{ du2021quantum, li2021quantum, abbas2021power}. 
Quantum variational autoencoders (QVAEs) had been primarily used for quantum state compression on adiabatic quantum computers \cite{romero2017quantum, khoshaman2018quantum} but rarely on gate-model quantum computers. When being implemented, quantum VAE/AE is more challenging than the usual quantum neural network (QNN) since it contains two major components, namely encoder and decoder. The scaling to high-dimensional dataset becomes even more challenging given limited number of qubits and the input-output mapping constraint imposed by measurements interpreted as the output. The output dimension is constrained to be less than the qubit count when measurement expectation value is taken as output.
We aim to address these scaling difficulties of quantum autoencoders by exploring various architectural techniques for drug discovery.

The actual QVAE architecture primarily depends on the dimension and scale of training dataset. A fully quantum autoencoder can achieve quantum learning advantages for normalized low dimension dataset, whereas a hybrid quantum classical model is necessary to implement the high-dimensional and large-scale dataset. Appropriate choices of quantum embedding methods also impact the quality of uploading classical data into quantum states in a Hilbert space.
Besides, the novel QNN architecture with patched quantum circuits, as shown in Fig. \ref{architecture}(b), is proposed to alleviate such dilemma. The strategy of adopting heterogeneous learning rates for quantum and classical layers is explored for counteracting the parameter range imbalance between the two layer types. Appropriate quantum layer depth is explored to improve learning quality as well. A suitable combination among these architectural factors is identified for learning different dimensional datasets accordingly.

We propose a new qubit-efficient scalable quantum generative autoencoder (SQ-VAE) to learn latent molecular distribution with reconstruction and sampling fidelity on par with classical ones. The architecture of classical and quantum VAEs are shown in Fig. \ref{architecture}. Extensive experiments on the architectural techniques are performed using molecular datasets with different dimensions including 8x8 QM9 \cite{qm9} and 32x32 PDBbind ligand \cite{pdb}. Additionally, two image datasets, 8x8 Digits and 32x32 grayscale CIFAR-10 \cite{krizhevsky2009learning}, are used to better visualize the learning qualities from classical and quantum autoencoders.

\textbf{Contributions}: We, (i) develop a scalable gate-model quantum generative autoencoder for reconstructing and sampling real-size drug molecules with better drug properties; (ii) demonstrate quantum learning advantages in terms of training speed and representation quality experimentally on the low-dimensional digit and molecule datasets; (iii) propose several novel architectural techniques including the patched quantum circuit and heterogeneous learning rates to learn high-dimensional data distributions.

\section{Preliminaries}
\label{sec:preliminaries}

\subsection{Computational Drug Discovery}
\label{sec:drug}

\begin{figure}
\centering
\includegraphics[width=7cm]{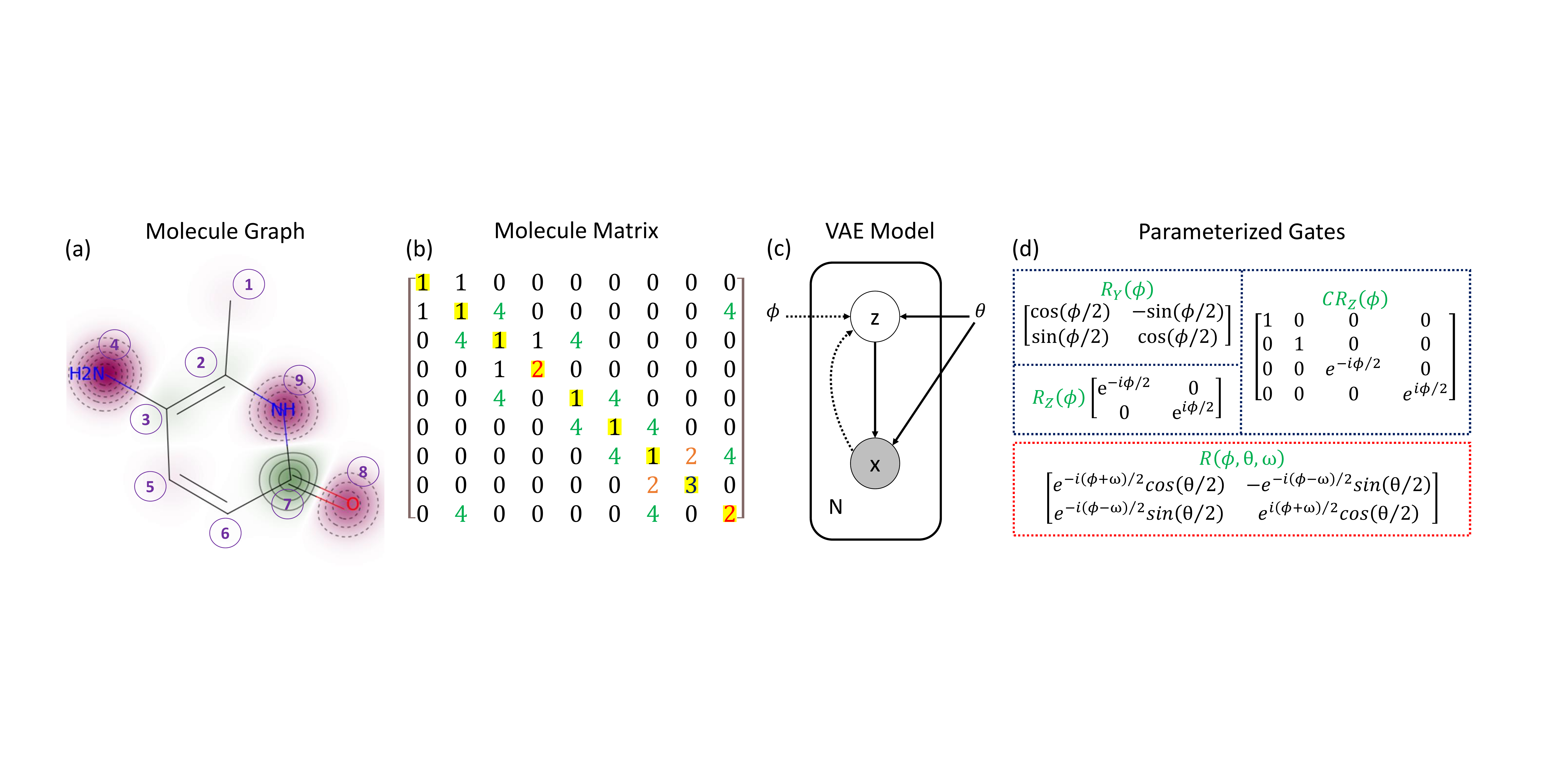}

\vspace{-3mm}
\caption{(a) A sample molecular graph where each circled number denotes the atom sequence; (b) the corresponding molecule matrix with diagonal elements representing atom types and off-diagonal ones representing bond types.}
\label{preliminaries}
\vspace{-3mm}
\end{figure}

A variety of classical computational approaches using GANs and VAEs have been explored for learning molecular distribution and generating new molecules from the learned chemical space \cite{molgan, kusner2017grammar}. In the present study, VAE is primarily selected because it learns both the inference network and generator network, the former of which helps with the instance-level matching between receptor protein and ligand. 
Quantum generative models have also been recently studied for leveraging the strong expressive power and intrinsic probabilistic nature of quantum circuits \cite{li2021drug, li2021quantum}. However, only small molecule discovery was targeted in these studies. 
One small molecular graph from QM9 and its matrix representation are shown in Fig. \ref{preliminaries}. Each diagonal element in the molecule matrix denotes the encoded atom type (1-C, 2-N, 3-O), and each off-diagonal element represents the encoded bond type (0-NONE, 1-SINGLE, 2-DOUBLE, 4-AROMATIC) in the molecular graph. Only heavy atoms excluding Hydrogen are displayed in the matrix.

\subsection{Variational Autoencoder}

Autoencoder (AE) consists of an encoder network, which converts an input vector to a code vector, and a decoder network which reconstructs the original input with high fidelity \cite{hinton1994autoencoders}. Its generative version of VAE additionally learns a variational model for the latent variables to capture the underlying sample distribution at the expense of less accurate reconstruction. Mathematically, VAE is represented by an inference network (a.k.a. encoder) $q_{\vphi}(\vz|\vx)$ and a generator network (a.k.a. decoder) $p_{\vtheta}(\vx|\vz)$ where $\vz$ denotes the latent variables \cite{vae}. The inference network represented by a neural network outputs parameters for learning the Gaussian distribution $q_{\vphi}(\vz|\vx) = \mathcal{N}(\mu_{\phi}(\vx), \vSigma_{\vphi}(\vx))$. The intractable true posterior $p_{\vtheta}(\vz|\vx)$ is approximated by the encoder network.
The generative model depends on a continuous variable $\vz$ and the posterior. Assuming a training set $S=\{\vx_i\}_{i=1}^N$, the parameters $\vphi$ and $\vtheta$ are jointly learned by minimizing the negative of evidence lower bound (ELBO) as follows:
\begin{align*}
&\min_{\vtheta, \vphi} \mathcal{L}_{\text{ELBO}}(\vtheta, \vphi; \vx) = \\ &-\mathbb{E}_{\vz \sim q_{\phi}(\vz|\vx)} [\log {(p_{\vtheta}(\vx|\vz))}]
+ D_{KL}[q_{\vphi}(\vz|\vx)||p(\vz)]
\end{align*}

\noindent where $p(z)$ is the prior assumed to be $\mathcal{N}(0, I)$. The first term $-\mathbb{E}_{\vz \sim q_{\phi}(\vz|\vx)} [\log {(p_{\vtheta}(\vx|\vz))}]$ in above equation is known as reconstruction loss, and the second term $D_{KL}(q||p)$ is the Kullback-Leibler (KL) divergence between the approximation density $q_{\vphi}(\vz|\vx)$ and the Gaussian prior. Continuous random noise is drawn from $p(\vz)$ for further approximating sampling new data points from $p_{\text{data}}(\vx)$.

\subsection{Quantum Neural Networks}

\textbf{Quantum gates} are quantum logic operations on a small number of qubits. A single-qubit operation behaves like a matrix-vector multiplication, and two-qubit operation are employed for creating quantum entanglement. Rotation gates are adopted for variational learning, and each rotation gate contains three trainable parameters.
Moreover, non-parametric CNOT gate (see Fig. \ref{architecture}(b)) is adopted for creating entanglement, and Pauli-Z gate for measuring the expectation value for each qubit.

\textbf{Quantum embedding} uploads a given classical data sample $x$ into its corresponding quantum state $\ket{\psi_x}$ in a Hilbert space. Embedding is a crucial part of quantum circuit which affects the computational power. 
For the vector $\vx \in \mathcal{R}^{d}$, the amplitude encoded state is defined as $\ket{x}=\frac{1}{\|\vx\|_2}\sum_{j=1}^{d}{x_j\ket{j}}$. Angle embedding maps the feature vector $\vx$ to the angles $\phi(\vx)=(\phi_1(x_1), \phi_2(x_2), \ldots, \phi_d(x_d))$ with which $d$ qubits have to be rotated to reflect each feature value $x_i$.
Amplitude embedding has the stringent input-output constraint, whereas angle embedding is not qubit efficient since it requires one qubit per feature.

\textbf{Measurement} provides classical information in terms of expectation, variance, or probabilities from quantum system. For expectations measured on the state $\ket{\psi}=a\ket{0}+b\ket{1}$, the average of the observable $\textit{Z}=$
$\big(\begin{smallmatrix}
  1 & 0\\
  0 & -1
\end{smallmatrix}\big)$ 
under the state mathematically refers to the following:
\begin{align*}
    \braket{\textit{Z}}_{\ket{\psi}} \equiv \bra{\psi}\textit{Z}\ket{\psi} \equiv \text{Tr}[\ket{\psi}\bra{\psi}\textit{Z}] = |a|^2-|b|^2 \in [-1, 1]
\end{align*}

\noindent where the expectation bound $[-1, 1]$ is obtained since the squared norm of amplitude represents a probability. For probabilities measured on the state, an array containing the probability $|\braket{i|\psi}|^2$ for each computational basis state $i$ is returned, satisfying $\sum_i|\braket{i|\psi}|^2=1$. 

\begin{figure*}
\centering
\includegraphics[width=17.5cm]{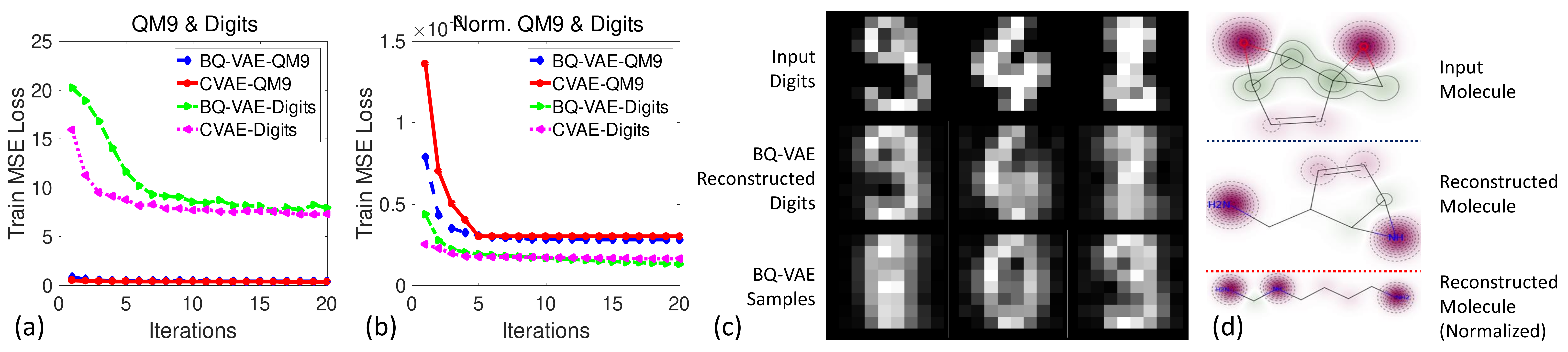}

\caption{(a) Comparison of train MSE losses between VAE and BQ-VAE on original Digits QM9 molecules; (b) comparison of train MSE losses on normalized Digits and QM9 molecules by dividing the respective L1 norm for all data points; (c) three sample digits, their respective BQ-VAE reconstructions, and three random samples from the learned generator network; (d) one sample QM9 molecule and its BQ-VAE reconstructions from input (original) and L1 normalized molecule, respectively.}
\label{bq-vae}
\vspace{-3mm}
\end{figure*}

\section{Variational Quantum Autoencoder}
\label{sec:qvae}

\subsection{Architectures}

The general QNN architecture consists of quantum embedding, repeatable variational entanglement layers, and measurement for constructing quantum generative and vanilla autoencoders. However, the specific choices of quantum embedding and measurement largely depend on the dimensions and scales of given dataset. Fully quantum autoencoders only apply to normalized-scale reconstruction since both expectation and probabilities have the upper bound of 1. High-dimensional data reconstruction would require the patched architecture to avoid the tradeoff between qubit-efficiency and input-to-output mapping constraint. For simplicity, the repeatable hidden layer is fixed with rotation gates $R(\psi, \theta, \omega)$ acting on each qubit, followed by CNOT gates with a periodic layout, as shown in Fig. \ref{architecture}(b). We demonstrate two quantum autoencoder architectures, \textit{i.e.}, baseline quantum VAE/AE (BQ-VAE/AE) and scalable quantum VAE/AE (SQ-VAE/AE) to learn the low-dimensional QM9 and high-dimensional PDBbind molecular datasets, respectively.

\subsection{Baseline Quantum Autoencoders}
\label{sec:bq}

BQ-VAE/AE adopts amplitude embedding and expectation output for encoder, and angle embedding and probability output for decoder. Low resolution 8x8 Digits set is used to compare BQ-VAE/AE performance relative to the classical counterpart. To demonstrate fully baseline quantum VAEs (F-BQ-VAEs), QM9 molecules and image digits are first normalized by directly dividing each non-negative feature value by their sum. Latent space dimension (LSD) is $log_2(64)=6$ by measuring expectation of each qubit. The LSD vector from the latent space is reconstructed into $2^6=64$ dimension by computing the probability of basis states. For original-scale (non-normalized) data points, a final fully connected (FC) classical layer is used to reconstruct feature values, resulting in the hybrid quantum variant (H-BQ-VAE/AE). Quantum hidden layers are repeated for $L=3$ times in BQ-VAEs/AEs. Similarly, classical encoder also takes 3 hidden linear layers followed by ReLU activation for reducing the dimensions to 32, 16, and 6, respectively, while classical decoder converts the dimensions in a reversed order. Table \ref{bqvae-weights} compares of number of trainable parameters between classical and baseline quantum autoencoders.

\begin{table}[h!]
\caption{Comparison of number of trainable parameters.}
\centering
\begin{tabular}{c c c c c}
 \hline
 Parameter Type & VAE(AE) & F-BQ-VAE(AE) & H-BQ-VAE(AE)\\ [0.5ex]
 \hline
 Quantum & 0 (0) & 108 (108) & 108 (108)  \\
 Classical & 5694 (5610) & 84 (0)  & 4286 (4202)  \\
 Total & 5694 (5610) & 192 (108) & 4394 (4310) \\
 \hline
 
\end{tabular}
\label{bqvae-weights}
\end{table}

Large number of classical parameters in H-BQ-VAE is due to the final linear layer with a [64, 64] transformation. Apart from using fewer parameters, BQ-VAE/AE even learns faster for normalized QM9 molecules in terms of the number of training epochs as shown in Fig. \ref{bq-vae}(b). However, Fig. \ref{bq-vae}(a) does not show quantum computational advantage for original-scale dataset due to the probability output constraint. Three normalized input samples from Digits dataset, their respective BQ-VAE reconstructions, and three randomly sampled digits from the learned generator network are shown in Fig. \ref{bq-vae}(c). Note that, normalized-scale digit images have the same quality as original-scale ones when being visualized. However, molecule reconstructions from original and normalized inputs differ so much that the latter hardly shares characteristics with the input molecule, as shown in Fig. \ref{bq-vae}(d). This indicates that normalization is not the feasible quantum autoencoder approach for learning QM9 molecular representation and high-dimensional PDBbind molecules. 

\subsection{Scalable Quantum Autoencoders}
\label{sec:sq}

\begin{figure}
\centering
\includegraphics[width=8cm]{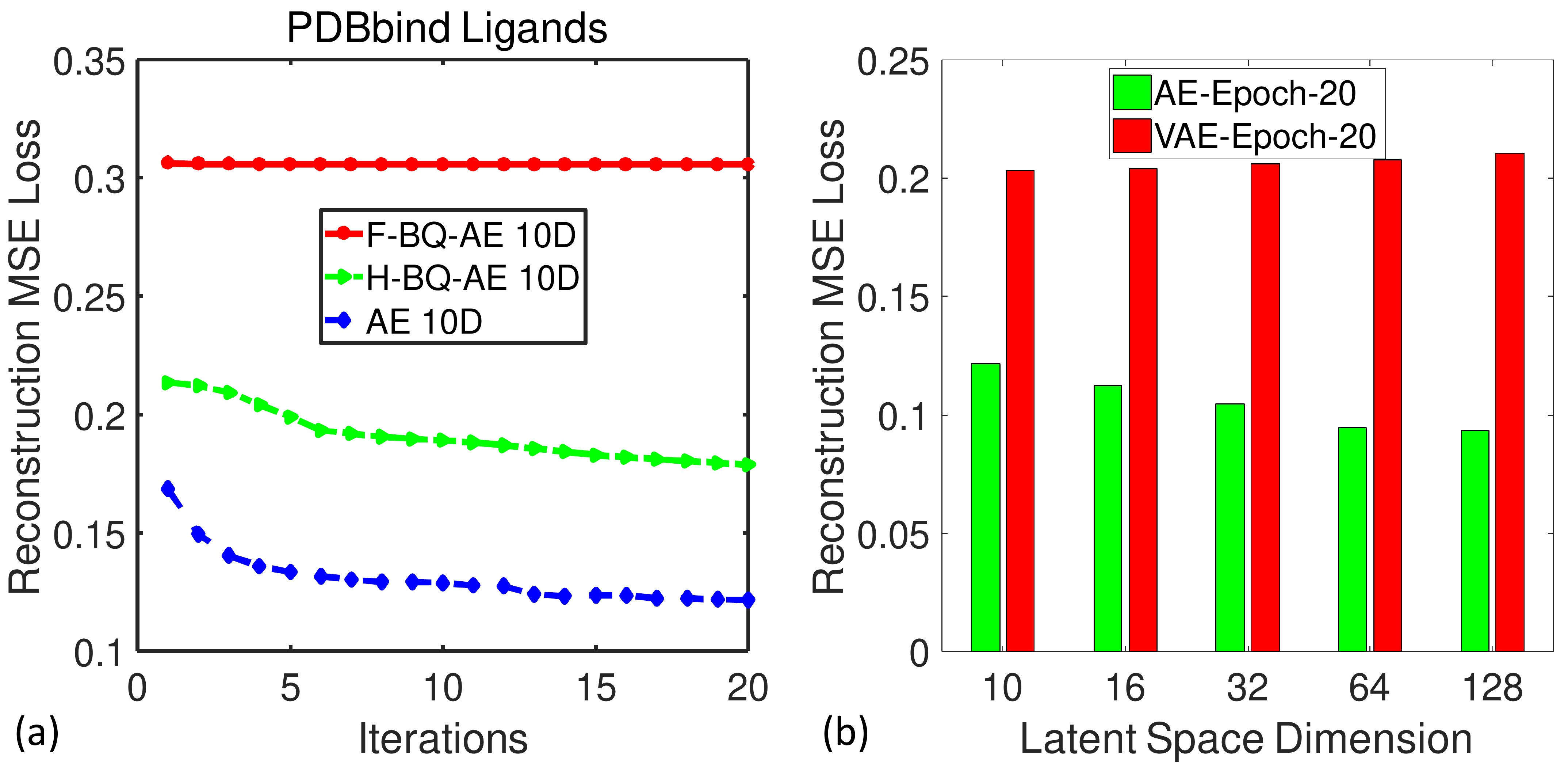}
\vspace{-3mm}
\caption{(a) Two BQ-AEs and the classical AE with three hidden layers applied to high-dimensional PDBbind ligands; (b) AEs with increasingly large latent space show lower test losses while VAEs remain almost unchanged.}
\label{bq-vae-lig}
\vspace{-4mm}
\end{figure}

Scalable quantum autoencoders including SQ-VAEs and SQ-AEs are developed for near-term quantum computers to learn high-dimensional dataset such as, PDBbind ligands filtered with up to 32 heavy molecules.
As illustrated in Fig. \ref{bq-vae-lig}, direct application of baseline quantum autoencoders to the 1024-dimensional (32x32) data points do not work well due to the stringent input-to-output mapping, i.e., latent space being too compact ($10=\log_{2}1024$ dimensional) to faithfully represent PDBbind ligands. The fully quantum variant, F-BQ-AE, hardly learns because probability values cannot be matched to the original-scale ligand matrices. The extra classical layer in the hybrid quantum model (H-BQ-AE) functions as the transformation to original scales. Fig. \ref{bq-vae-lig} (b) shows lower test losses for classical AEs with larger latent space dimensions whereas VAEs almost remain unchanged. Scalable quantum autoencoder relies on the patched quantum circuits for scaling to high-dimensional dataset given small size quantum computers. 
Angle embedding is used to encode the latent variable $\vz$ into the corresponding quantum state 
and measurement expectation value is returned as the output since the probabilities from 1024 basis states are too miniscule to be reconstructed.


Patched quantum circuit (see Fig. \ref{architecture}(b) is derived from the patch quantum GAN \cite{pan} which takes all features into each of the multiple circuits to obtain richer representation of fake digits. However, we partition the entire feature vector into multiple equal-sized sub-vectors, and each sub-vector is fed into a quantum sub-circuit. Therefore, our revised method occupies fewer qubits in each sub-circuit by embedding only a portion of the feature vector. Moreover, patched quantum circuit increases the output dimension by taking more expectation values as output. This alleviates the issue of having too compact latent space from directly using baseline quantum autoencoder. It is worth mentioning that the architecture advantage in terms of model complexity of SQ-VAE/AE over its classical counterpart becomes more pronounced, relative to the comparison aforementioned in Table \ref{bqvae-weights}, because of the higher dimensional input. The advantage here is not re-tabulated for brevity.

Aside from the patched circuit, we also present three following architecture-related factors that could affect SQ-VAE/AE performance. Firstly, the number of quantum hidden layers $L$ determines the number of trainable weights which relate to the trainability and expressiveness of variational quantum circuits. A suitable circuit depth is selected after a series of sensitivity studies. Secondly, similar to Fig. \ref{bq-vae-lig}(b), the learning quality of SQ-VAE/AE also depends on latent space dimension which in turn affects the number of patches in the circuit. 
Thirdly, quantum parameters fall in the range $[-\pi, \pi]$, whereas classical parameter space is much more vast given equal number of parameters. The small-scale of quantum parameters may require a different quantum learning rate schedule from classical one. This assumption is validated by adopting heterogeneous learning rates for quantum and classical layers.

\section{Experiments}
\label{sec:experiments}

This section discusses the experiment results of SQ-VAE/AE for learning high-dimensional molecules since the learning result of baseline quantum autoencoders has already been demonstrated on QM9 molecules in Section \ref{sec:bq}. 

\subsection{Dataset and Metrics}

The refined PDBbind 2019 dataset \cite{pdb} contains 4852 protein-ligand complexes, whereas  only ligand molecules are used to demonstrate the SQ-VAE/AE. We filter out molecules with more than 32 heavy atoms since 32x32 (a power of 2) dimension can directly leverage quantum information processing. Similar to QM9 dataset (see the 9x9 example in Fig. \ref{preliminaries}), PDBbind ligands are also represented by molecule matrices where diagonal elements indicate the encoded atom types (1-C, 2-N, 3-O, 4-F, 5-S) and off-diagonal elements indicate the encoded bond types. Each molecule matrix is symmetric, therefore, half of the off-diagonal elements could be removed. Ligands consisting of other atom types are removed. Therefore, only 2492 ligands are finally taken for learning SQ-VAE/AE. The dataset is split into 85\% and 15\% for train and test sets, respectively. SQ-VAE/AE performance on PDBbind ligands is evaluated using the metric of MSE loss together with three molecule property metrics of quantitative estimate of druglikeness (QED), log octanol-water partition coefficient (logP) and synthetic accessibility (SA) retrieved using RDKit \cite{rdkit}. Additionally, 32x32 resolution gray-scaled CIFAR-10 \cite{krizhevsky2009learning} images are adopted for visualizing SQ-AE reconstruction.

\subsection{Implementation Details}

Patched quantum circuit is included to scale the SQ-VAE/AE network to high-dimensional ligands. Other architecture-related factors \textit{i.e.}, adjustable quantum layer depth and heterogeneous learning rates are finalized through sensitivity studies. Both quantum encoder and decoder are connected to a classical layer to map the measurement to original ligand features.

All scalable quantum autoencoder variants are trained with a mini-batch of 32 molecules using the Adam optimizer with default $\beta_1=0.9$ and $\beta_2=0.999$ on a single RTX 2080 Ti GPU and the PennyLane simulation platform for the quantum stage. The learning rate is set to 0.001 throughout 20 training epochs for tuning quantum layer depth. A set of data points is randomly selected from the test set to evaluate SQ-VAE/AE reconstructions. Sampling results from SQ-VAEs are examined by sampling Gaussian noise from learned latent spaces.

\begin{figure}
\centering
\includegraphics[width=7cm]{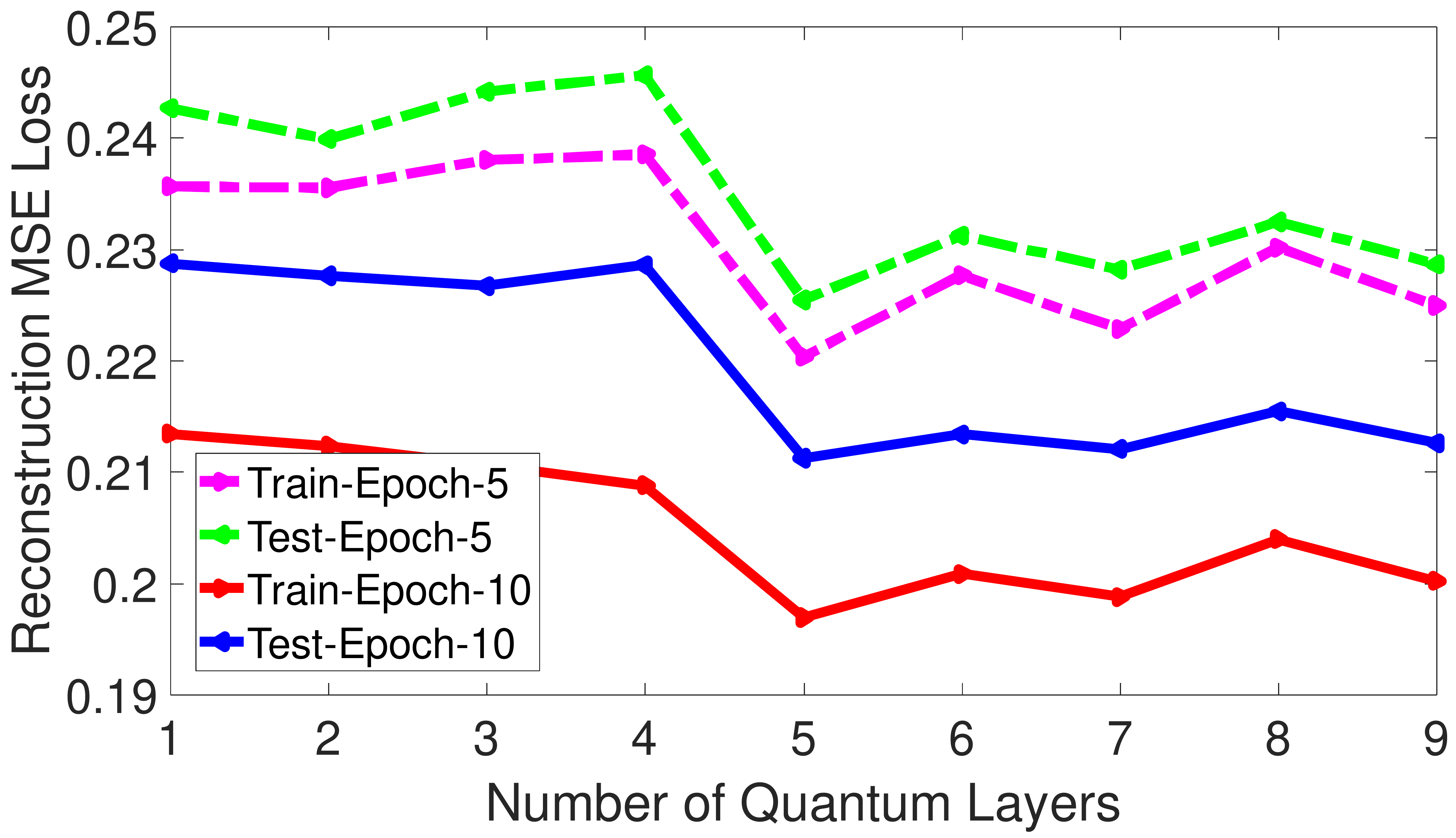}
\caption{Training and test reconstruction MSE loss comparison among SQ-VAEs with 1 to 9 strongly entangling layer(s).}
\label{qld}
\vspace{-3mm}
\end{figure}

\subsection{Ablation Study}

The number of spurious local minima depends exponentially on the number of quantum parameters \cite{you2021exponentially}. Therefore, the suitable quantum layer depth needs to be identified for better performance. 
Adopting heterogeneous learning rates for quantum and classical layers can speed up the overall learning.

The suitable quantum layer depth is identified by sweeping only the SQ-AE layers from 1 to 9. Note that SQ-VAE has the same architecture as SQ-AE except the extra latent variables. The losses after 5 and 10 epochs are shown in Fig. \ref{qld} where the configuration with 5 entangling layers produce the lowest test losses. The result indicates that SQ-AE with too few quantum layers hurts its expressive power, whereas too many layers possibly create unwanted number of spurious local minima. Thus the following SQ-VAE/AE experiments are all configured with 5 layers.

\begin{figure}
\centering
\includegraphics[width=7cm]{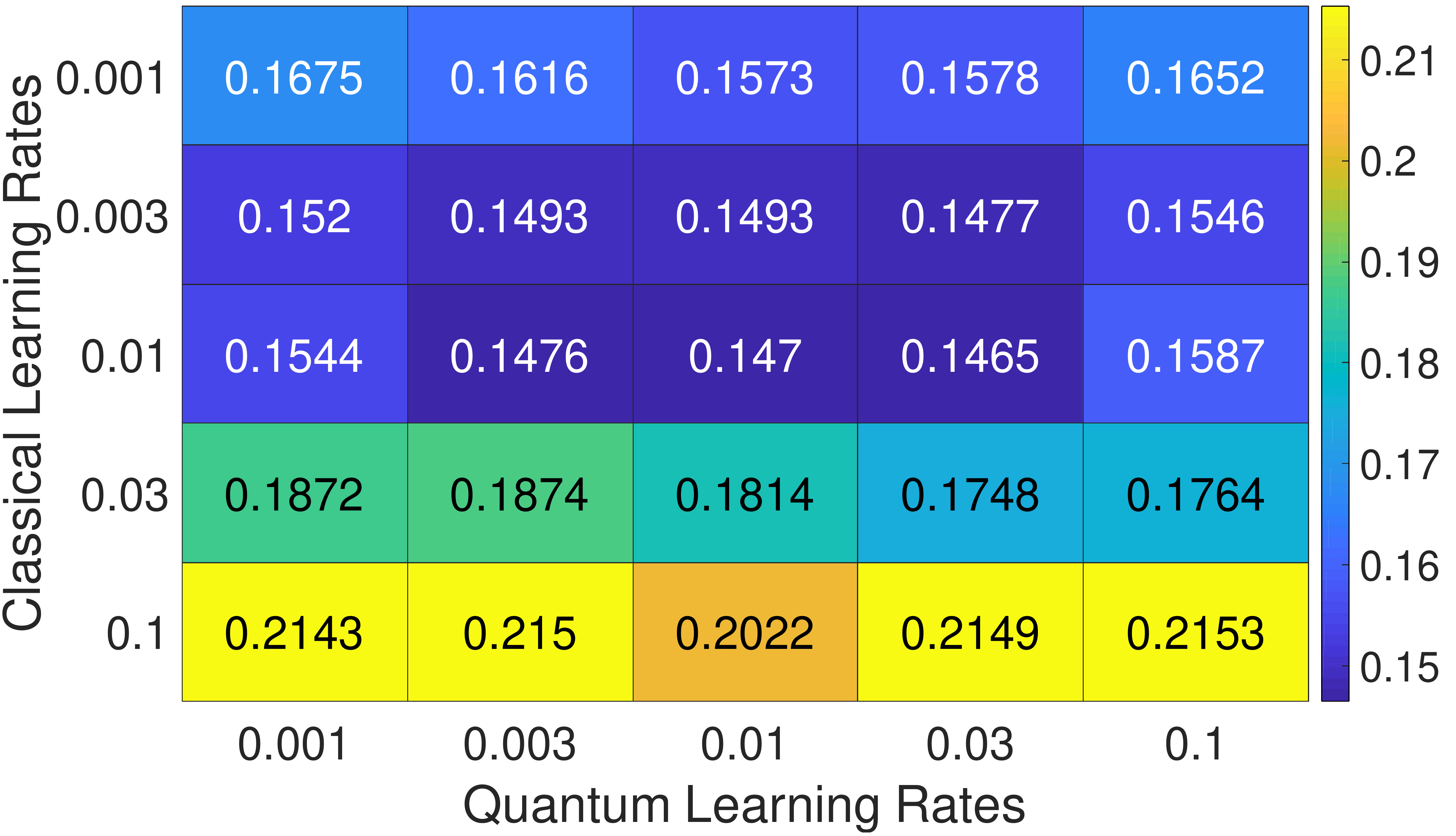}
\caption{Training loss comparison among SQ-VAEs configured with different combinations of quantum and classical learning rates.}
\label{hlr}
\vspace{-3mm}
\end{figure}

A single loss term is used to iteratively update the parameters in both quantum and classical layers. However, the quantum rotation angle parameter space is quite different from the classical parameter space. Therefore, studies of heterogenous learning rates are conducted only for SQ-AEs, due to architecture similarity with SQ-VAEs, to potentially balance the learning qualities in the hybrid algorithms. Five learning rates i.e., [0.001, 0.003, 0.01, 0.03, 0.1] are examined for both quantum and classical layers.
Fig. \ref{hlr} shows the sensitivity results of total 25 combinations. Classical learning rate of 0.01 and quantum learning rate of 0.03 produce the most favorable training loss. All the following experiments are configured with this combination.

\subsection{Evaluation Results}

\begin{figure*}
\centering
\includegraphics[width=17cm]{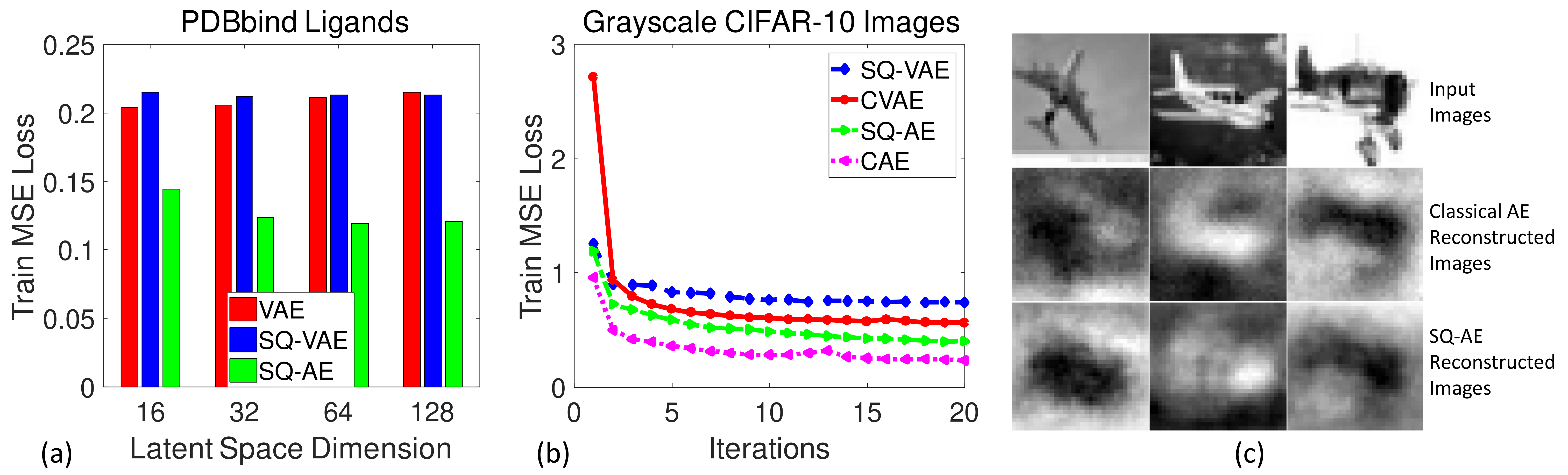}
\vspace{-3mm}
\caption{(a) Comparison of train losses between VAE and SQ-VAE/AE with different LSDs on PDBbind ligands; (b) comparison of train losses between VAE/AE and SQ-VAE/AE on grayscale CIFAR-10 images; (c) three sample CIFAR-10 images and their respective reconstructions from AE and SQ-AE.}
\label{sq-vae}
\vspace{-3mm}
\end{figure*}

As shown in Fig. \ref{bq-vae-lig}(a), baseline quantum autoencoders could not learn the high-dimensional PDBbind ligands well due to the input-output mapping constraint of the 10 LSD. Larger intermediate dimensions up to 64 improves the learning result for classical AEs (Fig. \ref{bq-vae-lig}(b)). However, input dimension and amplitude embedding determine the latent space dimension for quantum circuits. To increase LSD, patched quantum circuit in the scalable quantum autoencoder splits the holistic circuit into multiple patches. The performances of SQ-AEs with 4 different patches are evaluated. The variant with $p$ patches has $p\log_2 (1024/p)$ dimensions for latent space. Therefore, the LSD increases from 10 in baseline quantum autoencoders to 18, 32, 56 and 96 in scalable quantum autoencoders with 2, 4, 8 and 16 patches, respectively. Classical AEs with corresponding LSDs are created as well for comparison. To evaluate the sampling results using drug property metrics, SQ-VAE variants with corresponding patches are also conducted even though the negative MSE results are observed for SQ-VAEs with larger LSDs as shown in Fig. \ref{bq-vae-lig}(b).

Table \ref{sq-vae-results} compares the drug properties of 1000 new molecules generated from SQ-VAEs and VAEs with different LSDs after 20-epoch learning on PDBbind ligands. SQ-VAE with 18 LSD generates molecules with better logP and SA scores. The best QED score is achieved at SQ-VAE with 56 LSD using 8 circuit patches in the hybrid algorithm. VAEs with larger LSDs show slightly worse MSE losses as shown in Fig. \ref{sq-vae}(a), however, logP and SA scores increase considerably with increasing LSDs. SQ-VAE drug properties do not vary much with LSD.

\begin{table}[htbp]
\caption{Drug properties of sampled ligands from SQ-VAEs and VAEs with different LSDs.
}
\centering
\begin{tabular}{c c c c c c}
 \hline
 Metrics & LSD-18 & LSD-32 & LSD-56 & LSD-96 \\
 \hline
 \hline
VAE-QED & 0.138 & 0.179 & 0.139 & 0.142 \\
SQ-VAE-QED & 0.153 & 0.177 & \textbf{0.204} & 0.167 \\
 \hline
VAE-logP & 0.357 & 0.472 & 0.496 & 0.761 \\
SQ-VAE-logP & \textbf{0.780} & 0.616 & 0.709 & 0.740 \\
 \hline
VAE-SA & 0.192 & 0.292 & 0.307 & 0.599 \\
SQ-VAE-SA & \textbf{0.626} & 0.479 & 0.534 & 0.547 \\
 \hline
 
\end{tabular}
\label{sq-vae-results}
\end{table}

Sampling results can only be evaluated using generative autoencoders, while reconstruction can be evaluated using both generative and vanilla autoencoders. Reconstruction results from SQ-VAE/AE are approximately reflected by loss bars displayed in Fig. \ref{sq-vae}(a). Besides, reconstructions from autoencoders with 18 LSD are also evaluated using CIFAR images as shown in panels (b-c). SQ-VAE/AE shows reconstruction results on par with classical counterparts. Reconstructed images in panel (c) only show the sketches of original inputs for both classical and quantum AEs after learning in 20 epochs. The simulation inefficiency of quantum circuits precludes a longer learning period which leads to more accurate reconstruction.

\section{Conclusion}

This paper introduces a baseline quantum autoencoder that shows computational advantages for only low-dimensional molecules. The scalable quantum variants leverage the patched circuits and heterogeneous learning rates for scaling to high-dimensional ligands and show better sampling results in terms of drug properties and reconstruction results comparable to classical counterparts. The proposed scalable quantum autoencoder also applies to other tasks such as image generation.

{\small
\bibliographystyle{IEEEtran}
\bibliography{ref}

\begin{thebibliography}{10}
\providecommand{\url}[1]{#1}
\csname url@samestyle\endcsname
\providecommand{\newblock}{\relax}
\providecommand{\bibinfo}[2]{#2}
\providecommand{\BIBentrySTDinterwordspacing}{\spaceskip=0pt\relax}
\providecommand{\BIBentryALTinterwordstretchfactor}{4}
\providecommand{\BIBentryALTinterwordspacing}{\spaceskip=\fontdimen2\font plus
\BIBentryALTinterwordstretchfactor\fontdimen3\font minus
  \fontdimen4\font\relax}
\providecommand{\BIBforeignlanguage}[2]{{%
\expandafter\ifx\csname l@#1\endcsname\relax
\typeout{** WARNING: IEEEtran.bst: No hyphenation pattern has been}%
\typeout{** loaded for the language `#1'. Using the pattern for}%
\typeout{** the default language instead.}%
\else
\language=\csname l@#1\endcsname
\fi
#2}}
\providecommand{\BIBdecl}{\relax}
\BIBdecl

\bibitem{myers2001drug}
S.~Myers and A.~Baker, ``Drug discovery—an operating model for a new era,''
  \emph{Nature biotechnology}, vol.~19, no.~8, pp. 727--730, 2001.

\bibitem{li2018ligandomics}
W.~Li, I.-H. Pang, M.~T.~F. Pacheco, and H.~Tian, ``Ligandomics: a paradigm
  shift in biological drug discovery,'' \emph{Drug discovery today}, vol.~23,
  no.~3, pp. 636--643, 2018.

\bibitem{srinivasarao2017ligand}
M.~Srinivasarao and P.~S. Low, ``Ligand-targeted drug delivery,''
  \emph{Chemical reviews}, vol. 117, no.~19, pp. 12\,133--12\,164, 2017.

\bibitem{reymond2012exploring}
J.-L. Reymond and M.~Awale, ``Exploring chemical space for drug discovery using
  the chemical universe database,'' \emph{ACS chemical neuroscience}, vol.~3,
  no.~9, pp. 649--657, 2012.

\bibitem{ekins2019}
S.~Ekins, A.~C. Puhl, K.~M. Zorn, T.~R. Lane, D.~P. Russo, J.~J. Klein, A.~J.
  Hickey, and A.~M. Clark, ``Exploiting machine learning for end-to-end drug
  discovery and development,'' \emph{Nature materials}, vol.~18, no.~5, p. 435,
  2019.

\bibitem{vae}
D.~P. Kingma and M.~Welling, ``Auto-encoding variational bayes.'' in
  \emph{Proceedings of the International Conference on Learning Representations
  (ICLR)}, 2014.

\bibitem{gan}
I.~Goodfellow, J.~Pouget-Abadie, M.~Mirza, B.~Xu, D.~Warde-Farley, S.~Ozair,
  A.~Courville, and Y.~Bengio, ``Generative adversarial nets,'' in
  \emph{Advances in neural information processing systems}, 2014.

\bibitem{kusner2017grammar}
M.~J. Kusner, B.~Paige, and J.~M. Hern{\'a}ndez-Lobato, ``Grammar variational
  autoencoder,'' in \emph{International Conference on Machine Learning}.\hskip
  1em plus 0.5em minus 0.4em\relax PMLR, 2017, pp. 1945--1954.

\bibitem{molgan}
N.~De~Cao and T.~Kipf, ``Molgan: An implicit generative model for small
  molecular graphs,'' \emph{arXiv preprint arXiv:1805.11973}, 2018.

\bibitem{parmar2021dual}
G.~Parmar, D.~Li, K.~Lee, and Z.~Tu, ``Dual contradistinctive generative
  autoencoder,'' in \emph{Proceedings of the IEEE/CVF Conference on Computer
  Vision and Pattern Recognition}, 2021, pp. 823--832.

\bibitem{li2021drug}
J.~Li, M.~Alam, C.~M. Sha, J.~Wang, N.~V. Dokholyan, and S.~Ghosh, ``Drug
  discovery approaches using quantum machine learning,'' \emph{arXiv preprint
  arXiv:2104.00746}, 2021.

\bibitem{du2021quantum}
Y.~Du, M.-H. Hsieh, T.~Liu, D.~Tao, and N.~Liu, ``Quantum noise protects
  quantum classifiers against adversaries,'' \emph{Physical Review Research},
  vol.~3, no.~2, p. 023153, 2021.

\bibitem{li2021quantum}
J.~Li, R.~Topaloglu, and S.~Ghosh, ``Quantum generative models for small
  molecule drug discovery,'' \emph{arXiv preprint arXiv:2101.03438}, 2021.

\bibitem{abbas2021power}
A.~Abbas, D.~Sutter, C.~Zoufal, A.~Lucchi, A.~Figalli, and S.~Woerner, ``The
  power of quantum neural networks,'' \emph{Nature Computational Science},
  vol.~1, no.~6, pp. 403--409, 2021.

\bibitem{romero2017quantum}
J.~Romero, J.~P. Olson, and A.~Aspuru-Guzik, ``Quantum autoencoders for
  efficient compression of quantum data,'' \emph{Quantum Science and
  Technology}, vol.~2, no.~4, p. 045001, 2017.

\bibitem{khoshaman2018quantum}
A.~Khoshaman, W.~Vinci, B.~Denis, E.~Andriyash, H.~Sadeghi, and M.~H. Amin,
  ``Quantum variational autoencoder,'' \emph{Quantum Science and Technology},
  vol.~4, no.~1, p. 014001, 2018.

\bibitem{qm9}
R.~Ramakrishnan, P.~O. Dral, M.~Rupp, and O.~A. von Lilienfeld, ``Quantum
  chemistry structures and properties of 134 kilo molecules,'' \emph{Scientific
  Data}, vol.~1, 2014.

\bibitem{pdb}
Z.~Liu, Y.~Li, L.~Han, J.~Li, J.~Liu, Z.~Zhao, W.~Nie, Y.~Liu, and R.~Wang,
  ``Pdb-wide collection of binding data: current status of the pdbbind
  database,'' \emph{Bioinformatics}, vol.~31, no.~3, pp. 405--412, 2015.

\bibitem{krizhevsky2009learning}
A.~Krizhevsky, G.~Hinton \emph{et~al.}, ``Learning multiple layers of features
  from tiny images,'' 2009.

\bibitem{hinton1994autoencoders}
G.~E. Hinton and R.~S. Zemel, ``Autoencoders, minimum description length, and
  helmholtz free energy,'' \emph{Advances in neural information processing
  systems}, vol.~6, pp. 3--10, 1994.

\bibitem{pan}
H.-L. Huang, Y.~Du, M.~Gong, Y.~Zhao, Y.~Wu, C.~Wang, S.~Li, F.~Liang, J.~Lin,
  Y.~Xu \emph{et~al.}, ``Experimental quantum generative adversarial networks
  for image generation,'' \emph{arXiv preprint arXiv:2010.06201}, 2020.

\bibitem{rdkit}
``Rdkit: Open-source cheminformatics software,'' \emph{https://www.rdkit.org/}.

\bibitem{you2021exponentially}
X.~You and X.~Wu, ``Exponentially many local minima in quantum neural
  networks,'' in \emph{Proceedings of the 38th International Conference on
  Machine Learning}, 2021.

\end{thebibliography}
}

\end{document}